\documentclass[preprint,11pt,authoryear]{elsarticle}
\journal{Journal of the Mechanics and Physics of Solids}

\usepackage{color}

\usepackage{amsmath}
\usepackage{amssymb}
\usepackage{multirow}
\usepackage{enumitem}

\DeclareMathOperator{\sign}{sgn}

\newcommand{\be}[1]{\begin{equation}\label{#1}}
\newcommand{\ee}{\end{equation}}
\newcommand{\ba}[1]{\begin{eqnarray}\label{#1}}
\newcommand{\ea}{\end{eqnarray}}

\usepackage{booktabs}
\usepackage{color}
\usepackage{lineno,hyperref}
\modulolinenumbers[5]

\begin{document}
\begin{frontmatter}



\title{Flutter and divergence instability in the Pfl\"uger column: \\ experimental evidence of the Ziegler destabilization paradox}




\author[trento]{Davide Bigoni\corref{cor1}}
\address[trento]{University of Trento, DICAM, via Mesiano 77, 38123 Trento, Italy}
\cortext[cor1]{Corresponding author.}
\ead{davide.bigoni@unitn.it}

\author[newcastle]{Oleg N. Kirillov}
\address[newcastle]{Northumbria University, Newcastle upon Tyne, NE1 8ST, UK}

\author[trento]{Diego Misseroni}

\author[trieste]{Giovanni Noselli}
\address[trieste]{SISSA--International School for Advanced Studies, via Bonomea 265, 34136 Trieste, Italy}

\author[trento]{Mirko Tommasini}

%
%
%
\begin{abstract}

Flutter instability in elastic structures subject to follower load, the most important cases being the famous Beck's and Pfl\"uger's columns (two elastic rods in a cantilever configuration, with an additional concentrated mass at the end of the rod in the latter case), have attracted, and still attract, a thorough research interest. In this field, the most important issue is the validation of the model itself of follower force, a nonconservative action which was harshly criticized and never realized in practice for structures with diffused elasticity. An experimental setup to introduce follower tangential forces at the end of an elastic rod was designed, realized, validated, and tested, in which the follower action is produced by exploiting Coulomb friction on an element (a freely-rotating wheel) in sliding contact against a flat surface (realized by a conveyor belt). It is therefore shown that follower forces can be realized in practice and the first experimental evidence is given for both the flutter and divergence instabilities occurring in the Pfl\"uger's column. In particular, load thresholds for the two instabilities are measured and the detrimental effect of dissipation on the critical load for flutter is experimentally demonstrated, while a slight increase in load is found for the divergence instability. The presented approach to follower forces discloses new horizons for testing self-oscillating structures and for exploring and documenting dynamic instabilities possible when nonconservative loads are applied.

\end{abstract}


\end{frontmatter}

\section{Introduction}

A myriad of articles have been written after \cite{pfluger50,pfluger55}, \cite{ziegler_0,ziegler_1,ziegler_2}, \cite{beck}, \cite{bolotin}, \cite{herrmann_2} and \cite{L1987} have initiated the study of structures subject to follower tangential loads and, more generally, of mechanical systems under the action of non-conservative positional (circulatory) forces. Together with dissipative forces, circulatory forces constitute two major examples of nonconservative forces in physics \citep{kre,BS2016}. Especially intriguing and frequently counter-intuitive is the result of the interaction between the nonconservative forces and the gyroscopic and potential ones on stability of equilibria \citep{K2007,K2013,U2017}. This is a topic of great and growing interest not only in structural mechanics \citep{G1990}, but also in rotordynamics and gyrodynamics \citep{SBM2008}, robotics and automatic control \citep{TMS2011}, aeroelasticity \citep{P2017}, fluid-structure interactions \citep{MM2010}, smart materials \citep{KI2011}, biomechanics \citep{AET2013,RZRB2009}, cytoskeletal dynamics \citep{BD2016,DC2017} and even in astrophysics and geophysics \citep{C1984,KI2017}.

Flutter, a dynamical instability (a vibration motion of increasing amplitude) which passes undetected using static methods, is a phenomenon that can affect structures loaded with follower forces and occurs at a load much smaller than divergence instability (an exponentially growing motion). In fact, follower forces can in certain circumstances provide energy to the elastic system to which they are applied, so that a blowing-up oscillation is produced, which eventually leads to a periodic self-oscillation mode. In this context, stability is affected by dissipation in a detrimental way, so that even a vanishing small viscosity can produce a strong (and finite) decrease in the flutter load evaluated without keeping damping into consideration, a counter-intuitive effect known as the \lq Ziegler paradox' \citep{AY1974,bolotinzinzer,ziegler_0,B1956,bolotin,KV2010}.  It should also be noticed that, while the viscosity decreases the critical load for flutter, at the same time it {\it slightly increases} the load necessary to produce divergence instability\footnote{Research on divergence instability is less developed than that relative to flutter, because the imperative of traditional engineering was not to exceed critical loads. On the other hand, a recent approach to the mechanics of structures is their exploitation as compliant mechanisms for soft robotics or energy harvesting \citep{doare}, even in the range of large displacements and beyond critical loads. From this point of view it becomes useful to investigate phenomena occurring beyond the critical load for flutter instability, for instance divergence.}.

The most important problem emerged from the beginning of the research on flutter instability is the concept itself of a follower force, which has been often questioned and denied, up to the point that \cite{koiter} proposed the \lq elimination of the abstraction of follower forces as external loads from the physical and engineering literature on elastic stability' and claimed \lq beware of unrealistic follower forces'. 
If follower forces could not be realized in practice, all the mathematical models based on this concept and the effects theoretically predicted from these models, including dissipative instabilities and the Ziegler paradox, would become mere mathematical curiosities without any specific interest. 
For this reason, experiments were attempted from the very beginning and developed in two directions, namely, using water or air flowing from a nozzle \citep{herrmann,wood}, or using a solid motor rocket \citep{sugiyama,SKKR2000}, to produce a follower force at the end of an elastic rod or of a Ziegler double pendulum \citep{SW1975}. Neither of these methods was capable of exactly realizing a tangential follower force, in the former case because of hydrodynamical effects influencing the motion of the rod and in the latter because of the non-negligible and variable mass of the rocket, which burned so fast that a long-term analysis of the motion was prevented \citep{validsugi}. The fact that a follower, tangential force was never properly realized is clearly put in evidence in the thorough review by \cite{elishakoff}.

\begin{figure}[ht!]
  \begin{center}
    \includegraphics[width=0.95\textwidth]{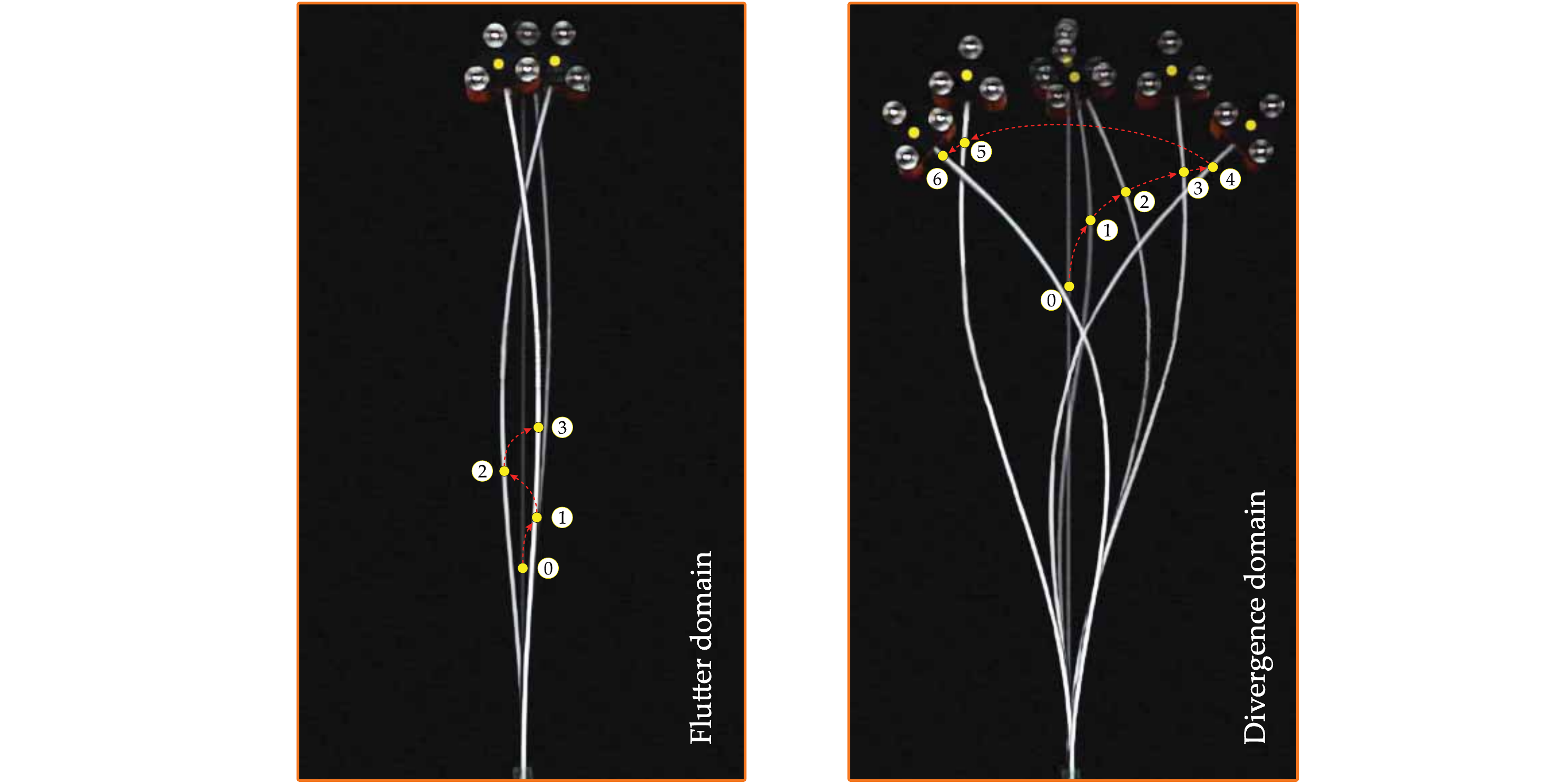}
    \caption{\footnotesize{Stroboscopic photographs (taken with a Sony PXW-FS5 camera at 240\,fps) of the initial motion of the Pfl\"uger's column when flutter instability (left) or divergence (right) occurs. The elastic rod corresponds to sample 5 of Table~\ref{table_beam}. A vertical load of 6.5\,N (40.0\,N) was applied for flutter (for divergence) and the conveyor belt was running at 0.1\,m/s. Note the different shape of the deformed rods at the beginning of the instability and their evolution towards a limit-cycle oscillation.}
}
    \label{fluutt}
  \end{center}
\end{figure}

A significant breakthrough was achieved by \cite{noselli}, who designed and tested a device capable of realizing a follower tangential force emerging at the contact with friction of a wheel constrained to slide against a moving surface. The device was successfully applied to the Ziegler double pendulum, but not to cases of diffuse elasticity, because difficulties related to early flexo-torsional buckling occurring in the structure prevented the use of the experimental setup. Therefore, the important cases of the Beck and Pfl\"uger's columns remained so far unexplored, such that the goals of the present study are:

\begin{itemize}

\item to extend the experimental investigation by \cite{noselli} to continuous elastic systems (in particular the Pfl\"uger's rod) through the design, realization and testing of a new mechanical apparatus to generate follower forces from friction. An example of the Pfl\"uger's rod during a flutter (left) and a divergence (right) experiment is shown in stroboscopic photos reported in Fig.~\ref{fluutt};

\item to provide another experimental evidence on the relation between Coulomb friction and dynamical instabilities occurring now in a deformable system with diffused elasticity;

\item to document (theoretically and experimentally) the existence of divergence instability occurring at loads higher than those for flutter for the Pfl\"uger's rod, an instability never addressed before;

\item to experimentally demonstrate that the introduction of damping lowers the critical load for flutter and increases the critical load for divergence (when this occurs at loads higher than those for flutter), thus verifying the concept of dissipation-induced instability and Ziegler's paradox \citep{kre,K2013};

\item to prove that, as a consequence of dissipation and nonlinearities, the systems, {\it both in flutter and in divergence conditions}, eventually attain a limit cycle in which a self-oscillating structure is obtained, in the sense explained by \cite{jenkins}.

\end{itemize}

The presented results: (i.)~show that concepts related to follower forces correspond to true physical phenomena, so that they do not represent a mere mathematical exercise, (ii.)~open new and unexpected possibilities to test the dynamical behaviour of structures under nonconservative loads (which include for instance the instability of a rocket subject to variable trust, the possibility to simulate aeroelastic effects without a wind tunnel, or to investigate flutter in models of protein filaments, or the interactions between mechanical components with frictional elements).

The paper is organized as follows: the design and realization of the experimental setup for follower forces is described in Section~\ref{flutter_machine}, while comparisons between experimental and theoretical results are presented in Section~\ref{exp_results}. The theoretical approach to the Pfl\"uger's column (and to the Beck's column as a particular case) and the identification of the parameters of the models of rods are deferred to~\ref{derivation_equation} and to~\ref{identification}, respectively. The theory of the modified logarithmic decrement approach used in the parameters identification procedure is reported in~\ref{app_A}.

Movies of the experiments can be found in the additional material available at \\ \url{www.ing.unitn.it/~bigoni/esm/Flutter_JMPS_final_Reduced.mp4}

\section{The \lq flutter machine': design, realization, and validation}\label{flutter_machine}

\begin{figure*}[ht!]
  \center
    \includegraphics[width=\textwidth]{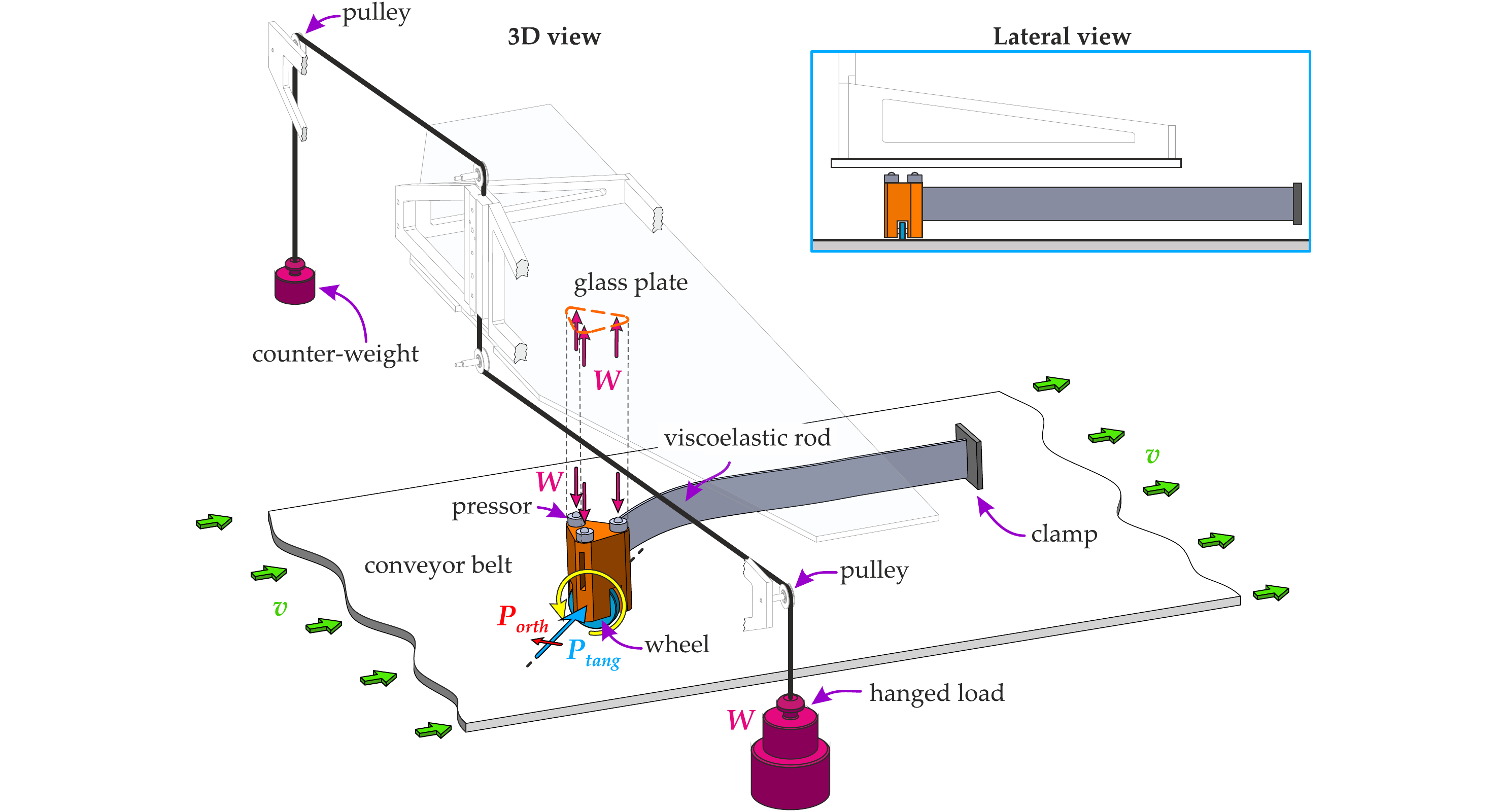}
    \caption{\footnotesize{The working principle of the flutter machine: a freely-rotating wheel is constrained to slide with friction against a moving substrate. The wheel is contained in a \lq loading head' endowed with a miniaturized load cell (used to measure the follower force) and a miniaturized accelerometer. The head is pressed vertically against the conveyor belt by a glass plate, indirectly loaded through a pulley system subject to a weight. A change in the applied weight leads, according to the law of Coulomb friction, to a change in the tangential force applied to the elastic rod.
    }}
    \label{fig2}
\end{figure*}

A new mechanical device, nicknamed \lq flutter machine', has been designed and realized to induce follower loads through friction at given points of an elastic system, which is now identified with either the Beck's or Pfl\"uger's column. The working principle of the device is sketched in Fig.~\ref{fig2} as a development of the idea by \cite{noselli} and is based on the friction that arises at the contact between a freely-rotating wheel and a moving surface. The wheel realizes a highly-anisotropic Coulomb friction, so that, in principle, only a force parallel to the wheel axis is produced ($P_{tang}$ in the figure). While the stroke of the device developed by \cite{noselli} was limited to about 1\,m, the new device realizes a virtually infinite stroke, because the sliding surface is obtained by means of a conveyor belt (C8N, from Robotunits). In this way, experiments of any time duration can be performed. A structure under test can be fixed at one end to a loading frame mounted across the moving belt, while subject at the other end to the follower force, which is transmitted to the structure through a \lq loading head' (realized in PLA with a 3D printer from GiMax 3D) endowed with the above-mentioned wheel (a ball bearing from Misumi-Europe). 
It should be noticed that the loading head is not exactly represented by a concentrated 
mass, because it possesses a non-null moment of inertia (so that a torque can also arise at the end of the rod in addition to the follower force). However, the head is made of PLA --a very light material-- and has a small size, so that 
it has been found (as will be reported in the following) that its rotatory inertia can be neglected, although the mass cannot be eliminated in practice, so that the Pfl\"uger's column, instead of the Beck's column, can be realized.

Measurement of the follower force is achieved by means of a miniaturized load cell (XFTC301 from TE connectivity) mounted inside the \lq loading head' together with a miniaturized accelerometer (352A24 from PCB Piezotronics).  The amount of the follower force can be calibrated as proportional, through Coulomb friction law, to a vertical load pressing the head against the conveyor belt ($W$ in the figure). This load is provided through contact of the head (on which three pressors are mounted to minimize friction, thus leaving the head free of moving) with a 5\,mm thick tempered glass plate, which is in turn loaded through a weight (a tank filled with water) indirectly applied by means of a double-pulley system. The pulley system is made in such a way that it can completely compensate for the weight of the loading frame, so that any vertical load can be applied starting from zero.

Note that the loading system of the rod's head is the most important design difference with respect to the earlier testing apparatus realized by \cite{noselli}. In fact, in the previous apparatus the amount of the follower load was controlled by exploiting the structure under testing as a lever, so that the setup was useful only for discrete structures (e.g., the Ziegler double pendulum) which, having an infinite torsional stiffness, do not display flexo-torsional instability. Hence, the design of the flutter machine was driven by the aim to overcome such limitation and extend the experimental study of flutter and divergence instability to structural systems with diffused elasticity.

\begin{figure}[ht!]
  \begin{center}
    \includegraphics[width=\textwidth]{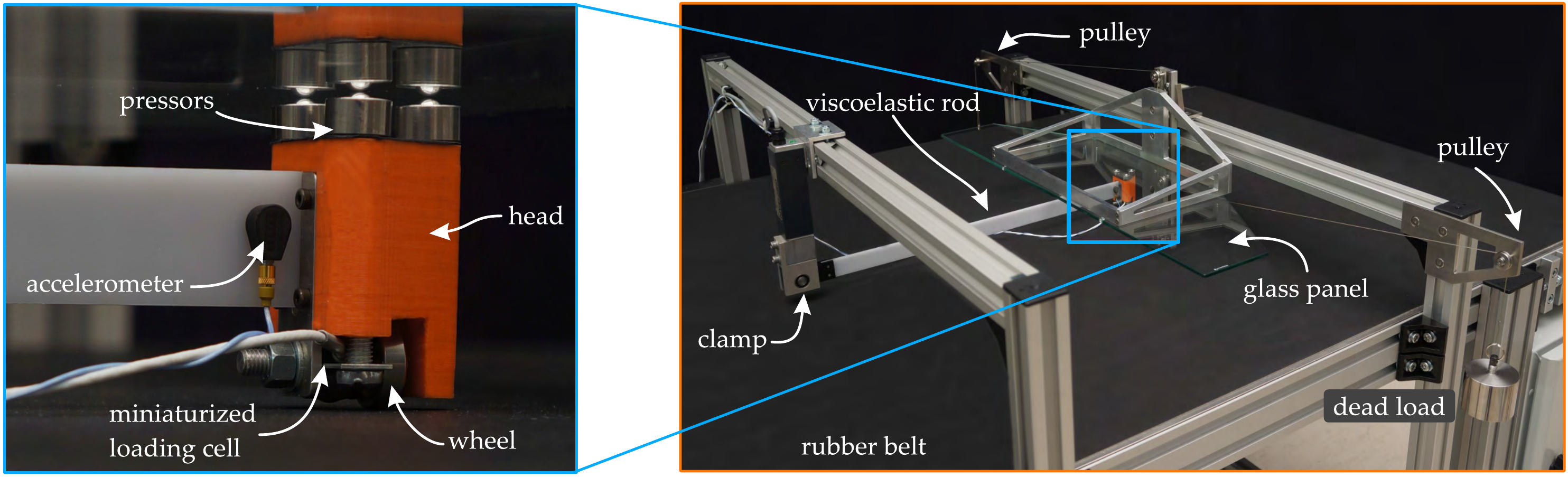}
    \caption{\footnotesize{A photograph of the \lq flutter machine' (on the right) showing the conveyor rubber belt and the polycarbonate rod connected at one end to a load cell and to the head transmitting the follower force at the other end. Modulation of the follower force is achieved by varying the vertical load acting on the head and transmitted through contact with a glass plate. Notice the double-pulley loading system. A detail is reported (on the left) of the head introducing the follower force to the end of the Pfl\"uger's column. The upper edge of  the head is in contact through three pressors with the glass plate transmitting the vertical load, while its lower edge contains the wheel sliding against the conveyor belt. The head contains a miniaturized load cell to measure the follower force, and a miniaturized accelerometer is mounted near the head.
}}
    \label{fig3_new}
  \end{center}
\end{figure}

Experimental results collected by exploiting the \lq flutter machine' will be presented in the following sections together with theoretical predictions. We anticipate here that all the data from the load cell and from the accelerometer were acquired with a NI cDaq-9172 system (from National Instruments) interfaced with LabVIEW 13. Also, the conveyor belt was actuated with a SEW-Eurodrive motor and controlled with an inverter (E800 from Eurodrive), so that its velocity could range between 0 and 0.3\,m/s. As regards the elastic rods that model the Pfl\"uger's column, these were realized in polycarbonate (white 2099 Makrolon UV from Bayer of Young's modulus $E\simeq$ 2344\,MPa, Poisson's ratio $\nu=$ 0.37 and volumetric mass density $m \simeq$ 1185\,kg\,m$^{-3}$). In order to measure the reaction force at the clamped extremity of the tested rods, this was endowed with a load cell (OC-K5U-C3 from Gefran). A photograph of the experimental setting  is reported in Fig.~\ref{fig3_new} together with a detailed view of the loading head.

Assuming perfect transmission of loads, perfect sliding with ideal Coulomb friction at the wheel contact, and null spurious frictional forces, a purely tangential follower load should be generated at the end of the Pfl\"uger's column. However, for practical reasons the flutter machine is not capable of perfectly realizing such an ideal condition and discrepancies can arise from different sources, the most important of which are listed and commented below.

\begin{enumerate}[label*=\arabic*.]

\item {\it Imperfect transmission of the vertical load $W$ at the head of the elastic rod.} To evaluate the performance of the load transmission system against the pressors mounted on the head of the elastic rod, a load cell has been placed between the glass plate and the belt. The load transmission mechanism was found to be almost perfect and insensitive to the position of the head on the glass plate, so that the mechanical components of the double-pulley loading system do not introduce any significant friction.

\item {\it Friction at the glass plate/pressors contact.} The pressors at the glass plate/head contact introduce a spurious friction in the system, that has been measured and found to be negligible.

\item {\it Generation of a spurious force orthogonal to the tangent to the rod at its end.} A departure from the pure tangentiality of the follower force is due to rolling friction and inertia of the wheel, and determines a non-null component of the follower load orthogonal to the tangent to the rod at its end, i.e. $P_{orth}$ in Figs.~\ref{fig2} and \ref{fig6}. This component, which adds to the tangential force $P_{tang}$, was directly measured with a load cell (XFTC301 from TE connectivity) using a rigid bar (instead of the deformable rod employed in the experiments) during sliding of the belt against the wheel. Different inclinations of the rod (15$^\circ$, 30$^\circ$, 45$^\circ$ and 60$^\circ$) were investigated for several applied vertical loads, see Fig.~\ref{fig6}. With these experiments the ratio between the orthogonal and tangential components of the load, namely, $v'(l)\bar{\chi}=\arctan(P_{orth}/P_{tang})$, where $\bar{\chi}=1-\chi$ and $v'(l)$ is the derivative of the rod's deflection at its end, was experimentally estimated. In particular, the mean value of $v'(l)\bar{\chi}=0.092$ was established.

\begin{figure}[ht!]
  \begin{center}
    \includegraphics[width=\textwidth]{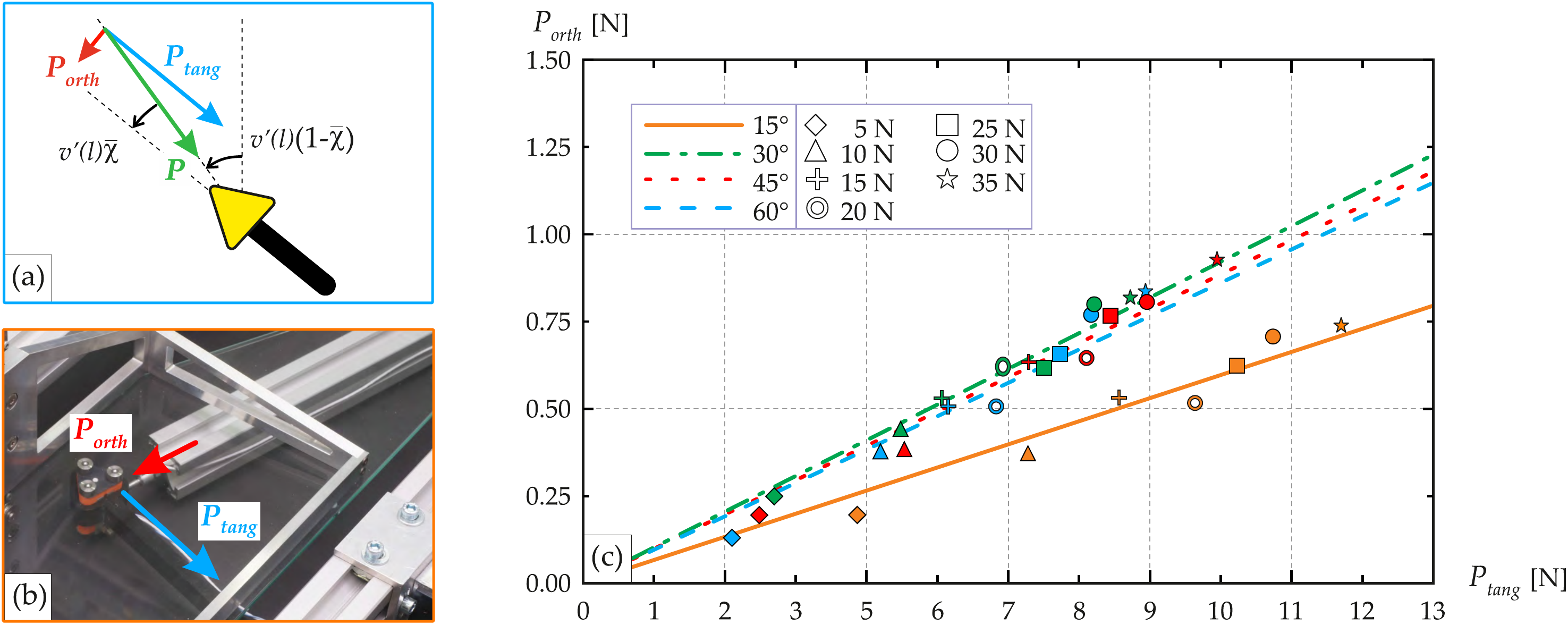}
       \caption{\footnotesize{(a)~A sketch of the tangential, $P_{tang}$, and orthogonal, $P_{orth}$, components of the follower force $P$ acting at the end of the rod. (b)~A photograph of the experimental setting exploited to measure the two components of the follower force $P$. Specifically, the tangential component is measured with the load cell inside the head, whereas the orthogonal component is measured with an external load cell. (c) Tangential and orthogonal components of the follower force $P$ acting on the loading head as measured for different vertical loads $W$ and different inclinations of the rod. Best fitting lines are reported for the distinct inclinations of the rod, leading to a mean value of $v'(l)\bar{\chi}=\arctan(P_{orth}/P_{tang})=$ 0.092.
}}
    \label{fig6}
  \end{center}
\end{figure}

\item {\it Deviation from the simple law of Coulomb friction at the sliding contact between the steel wheel and the rubber conveyor belt.} It has been experimentally determined that:

\begin{enumerate}[label*=\arabic*.]

\item the law of Coulomb friction at the wheel/conveyor belt contact holds only as a first approximation. In fact, the coefficient of friction was found to be a function of the vertical load applied to the head. The dependence of the friction coefficient upon the vertical load $W$ was investigated with a rigid rod replacing the viscoelastic rod. During experimentation, the velocity of the belt was kept constant at 0.1\,m/s. Experimental results are reported in Fig.~\ref{fig12}(a) together with their interpolation
\begin{equation}
\label{nonmirompere}
\mu_0(W)=-2.406+0.013W+(0.288+0.0039W)^{-1} ,
\end{equation}
which was used in the modelling. Notice that a nonlinearity of the friction coefficient at a rubber/steel contact (the belt is made of rubber) was previously documented in \cite{maegawa};
\begin{figure}[ht!]
  \begin{center}
    \includegraphics[width=\textwidth]{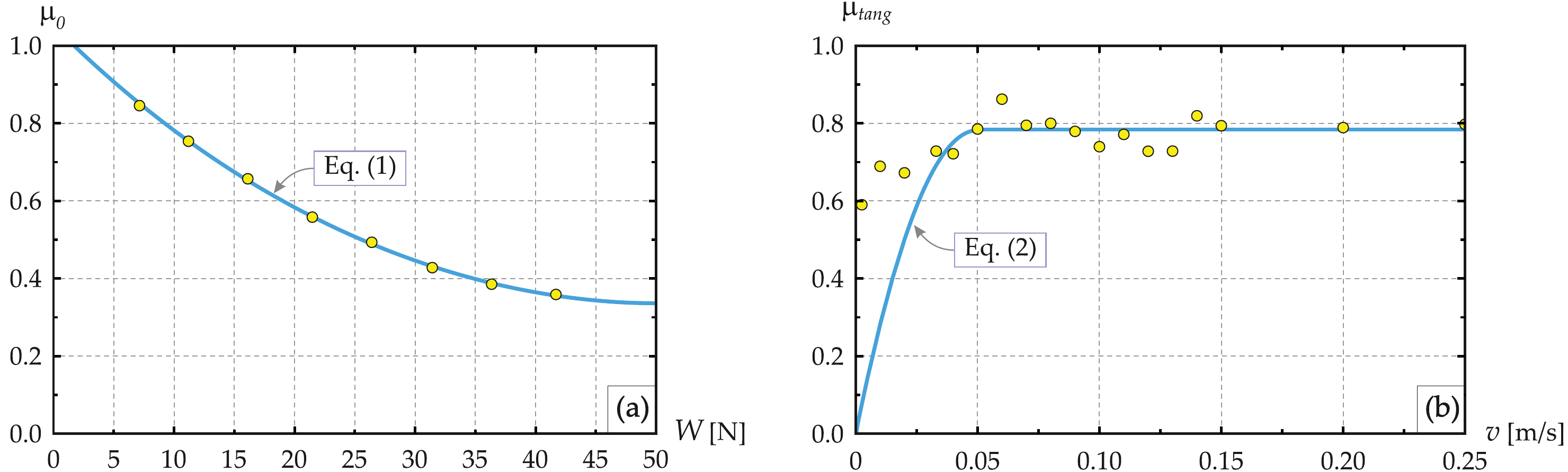}
    \caption{\footnotesize{(a)~Dependence of the dynamic friction coefficient $\mu_0$ at the wheel/belt contact on the applied vertical load $W$ for a fixed velocity of the belt, namely 0.1\,m/s. The experimental data (yellow spots) are fitted with the non-linear interpolation of Eq.~(\ref{nonmirompere}). (b)~Dependance of the dynamic friction coefficient $\mu_{tang}$ at the wheel/belt contact on the velocity of the belt for a fixed vertical load $W=10$\,N. The experimental data (yellow spots) follow the law proposed by \cite{oden} and \cite{martins}, that is Eq.~(\ref{martins}).
}}
    \label{fig12}
  \end{center}
\end{figure}

\item the friction coefficient is sensitive to the sliding velocity at the wheel/belt contact and exhibits a nonlinear behaviour characterized by an increase with velocity and later a stabilization, which occurs from approximately 0.05\,m/s. The dependence of the friction coefficient on the velocity was measured (at a constant vertical load of 10\,N) with the experimental setup used to address point (4.1), and found to follow the law provided by \cite{oden} and \cite{martins}. In particular, with reference to the friction coefficient $\mu_0(W)$ introduced in equation (\ref{nonmirompere}), the tangential coefficient of friction determining the relation between the follower force $P_{tang}$ and the weight $W$ as $P_{tang}=\mu_{tang} W$ can be written in the form
\be{martins}
\mu_{tang} =\mu_0(W)
\begin{cases}
\displaystyle \sign\left( v_{tang}\right) \,\,\,  \text{if}\,\,\, v_{tang} \not\in [-\varepsilon,\varepsilon] \\
\\
\displaystyle \left( 2 - \frac{\left| v_{tang} \right|}{\varepsilon}\right) \frac{v_{tang} }{\varepsilon} \,\,\,  \text{if}\,\,\, v_{tang} \in [-\varepsilon,\varepsilon] \\
\end{cases}
\ee
where $v_{tang}$ is the tangential component of the wheel/belt relative sliding velocity and $\varepsilon = 0.05$\,m/s is a parameter. The correspondence between experiments and the law of equation~(\ref{martins}) is reported in Fig.~\ref{fig12}(b). In view of the experimental results reported in Fig.~\ref{fig6}, the coefficient $\mu_{orth}$ is defined by the same equation (\ref{martins}),
but with a value of the dynamic friction coefficient equal to $0.09 \mu_0$ and with the velocity $v_{orth}$ replacing $v_{tang}$.

\end{enumerate}

\end{enumerate}

\section{Experimental results versus theoretical and computational predictions}\label{exp_results}

We provide here a detailed analysis of the results obtained by means of the experimental setting introduced in Section~\ref{flutter_machine}. While the onset of flutter and divergence instability can be accurately predicted via a theoretical study of the governing equations in their linearized form, a computational approach is needed to capture the nonlinear dynamics of the system at large displacements and rotations.

\subsection{Onset of flutter and divergence in the presence of internal and external dissipation}

The theoretical approach to the Beck's and Pfl\"uger's columns, when both the internal and external dissipations are present, is based on the analysis of an axially pre-stressed,
rectilinear and viscoelastic rod, characterized by the transversal deflection $v(x,t)$, function of the axial coordinate $x$ and of time $t$. A moment-curvature viscoelastic constitutive relation of the Kelvin-Voigt type is assumed in the form
\begin{equation}
\label{uno}
\mathcal{M}(x,t) = - EJ v''(x,t) - E^*J \dot{v}''(x,t) ,
\end{equation}
where a superimposed dot denotes the time derivative and a prime the derivative with respect to the coordinate $x$, $E$ and $E^*$ are the elastic and the viscous moduli, respectively, and  $J$ is the moment of inertia of the rod. The shear force $\mathcal{T}(x)$ can be computed as the derivative of the bending moment, so that for constant moduli $E$ and $E^*$, it can be written as
\begin{equation}
\mathcal{T}(x,t) = - EJ v'''(x,t) - E^*J \dot{v}'''(x,t) .
\end{equation}

\begin{figure}[ht!]
  \begin{center}
    \includegraphics[width=\textwidth]{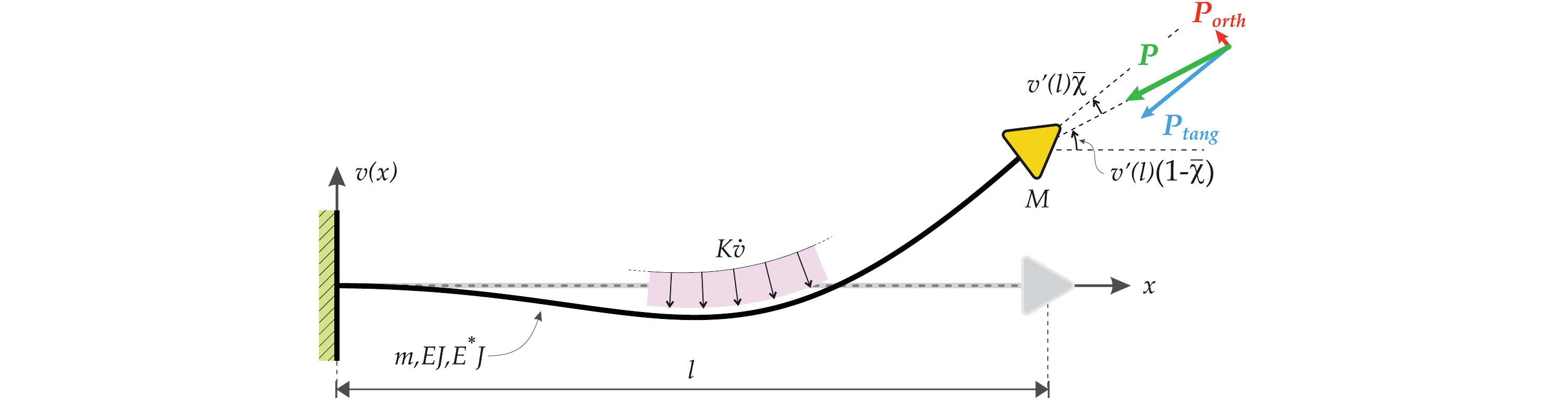}
    \caption{\footnotesize{Structural model for the Pfl\"uger's column: a viscoelastic cantilever rod (of length $l$, mass per unit length $m$, and elastic and viscous bending stiffnesses $EJ$ and $E^*J$, respectively) with a concentrated mass $M$ at its free end is loaded by a partially tangential follower load $P$.
}}
    \label{fig_bound_cond}
  \end{center}
\end{figure}

The linearized differential equation of motion which governs the dynamics of a rectilinear viscoelastic rod, loaded by an axial force $P$ (positive when compressive) and a transversal distributed load proportional through a coefficient $K$ to the velocity $\dot{v}$ (which models for instance the air drag) is
\be{eq_diff}
EJ v''''(x,t) + E^*J\dot{v}''''(x,t) + P v''(x,t) + K \dot{v}(x,t) + m \ddot{v}(x, t) = 0,
\ee
where $m$ is the mass density of the rod per unit length, see Fig.~\ref{fig_bound_cond}.

The solution to the Pfl\"uger's column problem (of which the Beck's column is a particular case) is obtained by imposing on the differential equation~\eqref{eq_diff} the following boundary conditions
\begin{equation}
\begin{array}{ll}
v(0, t) = v'(0, t) = 0        & \mbox{at the clamped end,} \\ [5 mm]
\mathcal{M}(l, t) = -J[E v''(l,t) - E^* \dot{v}''(l,t)] = 0 & \mbox{at the loaded end,} \\ [5 mm]
\mathcal{T}(l, t) = -J[E v'''(l,t) - E^* \dot{v}'''(l,t)] = P \bar\chi v'(l,t) - M \ddot{v}(l,t) & \mbox{at the loaded end,}
\end{array}
\label{contorno}
\end{equation}
where $\bar{\chi}$ measures the inclination of the applied tangential force $P$, such that $\bar{\chi}=0$ corresponds to a purely tangential follower load. Notice that the Beck's column problem is obtained by setting $M=0$. By introducing the dimensionless quantities
\be{dimensionless}
\xi =\frac{x}{l}, ~ \tau = \frac{t}{l^2}\sqrt{\frac{E J}{m}}, ~ p =\frac{P l^2}{E J}, ~\alpha =\arctan\left(\frac{M}{m l}\right), ~\eta = \frac{E^*}{E l^2}\sqrt{\frac{E J}{m}}, ~\gamma = \frac{K l^2}{\sqrt{m E J}} ,
\ee
the governing equation~\eqref{eq_diff} can be rewritten as
\begin{equation}
\label{checazzo}
v''''(\xi, \tau) + \eta \dot{v}''''(\xi,\tau) + p v''(\xi,\tau) + \gamma \dot{v}(\xi,\tau) + \ddot{v}(\xi,\tau) = 0 ,
\end{equation}
where now a prime denotes differentiation with respect to $\xi$ and a dot differentiation with respect to $\tau$. It is now expedient to assume time-harmonic vibrations of dimensionless pulsation $\omega$ as
\begin{equation}
\label{sclero}
v(\xi,\tau)=\tilde{v}(\xi) \exp(\omega \tau) ,
\end{equation}
such that equation~\eqref{checazzo} is transformed into a linear differential equation for $\tilde{v}(\xi)$, which can be solved in an explicit form. Specifically, imposition of the boundary conditions~\eqref{contorno}, transformed via equations~\eqref{dimensionless} and~\eqref{sclero}, yields an eigenvalue problem for the pulsation $\omega$, which can be solved for given values of:~(i.)~dimensionless load $p$, (ii.)~mass contrast $\alpha$, (iii.)~rod viscosity $\eta$, (iv.)~external damping $\gamma$ modelling the air drag, and (v.)~inclination of the load $\chi$. The details of this analysis are deferred to~\ref{derivation_equation}.

The stability of the structure can be judged on the basis of the nature of the pulsation $\omega$. In particular, the system is unstable when the real part of the pulsation is positive, Re[$\omega$] $>$ 0; in this case flutter instability occurs when $\omega$ is complex with non vanishing imaginary part, whereas the system becomes unstable by divergence when $\omega$ is purely real and positive. In view of this, flutter corresponds to a blowing-up oscillation, whereas divergence to an exponential growth.

\begin{figure}[ht!]
\begin{center}
\includegraphics[width=\textwidth]{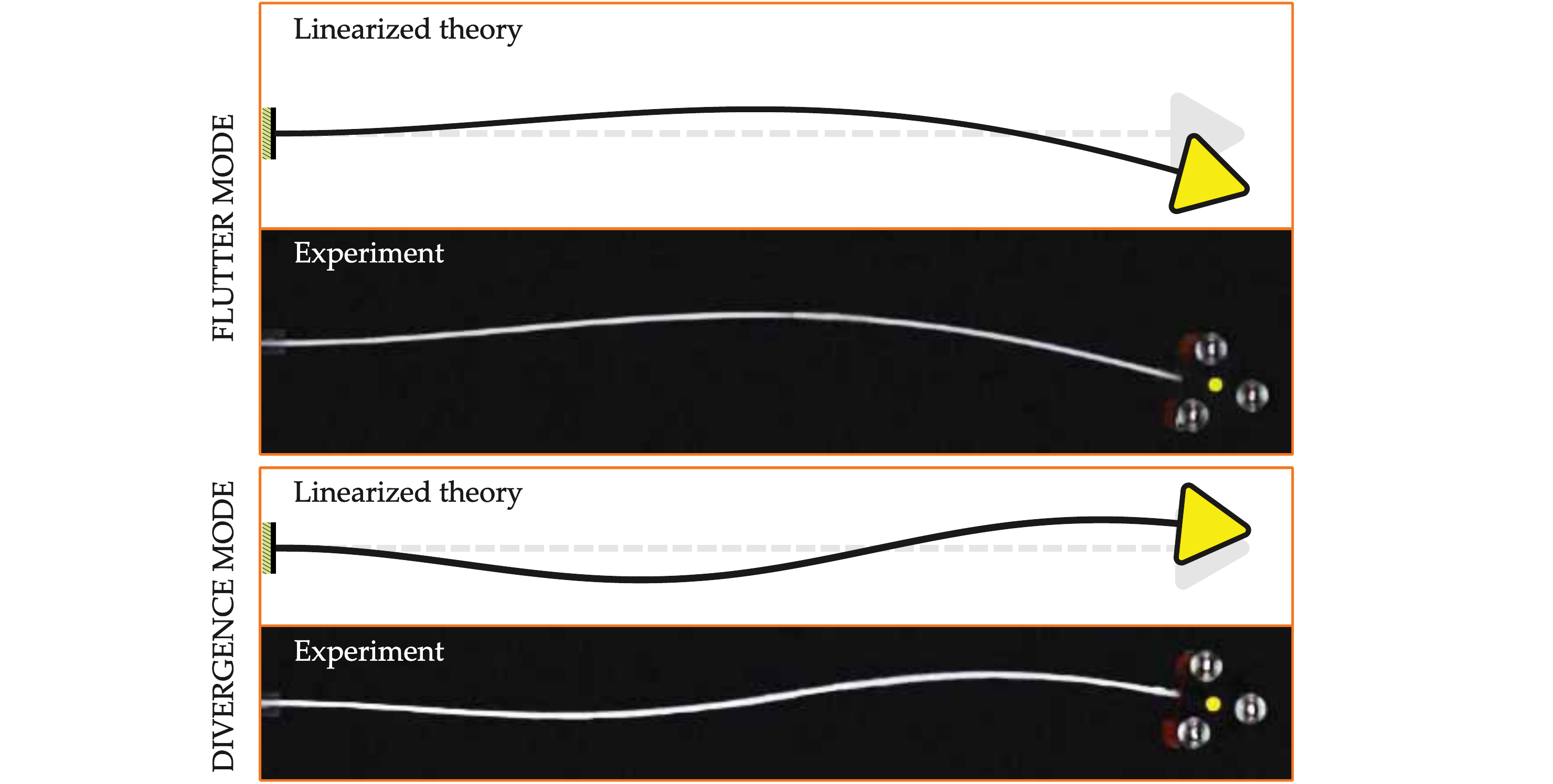}
\caption{\footnotesize{Eigenmodes associated to flutter instability (upper part) and divergence instability (lower part) compared with two photographs referred to two experiments performed on sample 5 of Table~\ref{table_beam}.
}}
\label{autovettori2}
\end{center}
\end{figure}

The eigenmodes associated to flutter instability and divergence instability can be found from the eigenvectors of matrix~(\ref{matriciaccia}), once the critical pulsation $\omega$ has been determined. As an example, referred to sample 5 of Table~\ref{table_beam}, two eigenmodes, one corresponding to flutter instability (6.5\,N of applied vertical load) and the other to divergence instability (40.0\,N of applied vertical load) are reported in Fig.~\ref{autovettori2}, together with two photos taken during experimentation.
Note the difference between the shapes of the two modes and the remarkable correspondence with the photos, which correspond to the onset of the two motions shown in Fig. \ref{fluutt}. The difference in the shape of the modes for flutter and for divergence has been used during the experiments to discriminate between the two instabilities (see the following discussion).

Dissipation has a complex effect on flutter and divergence instability \citep{K2013}. Without entering a long discussion (a review of this topic together with new results on the effects of external and internal damping has been recently provided by \cite{KV2010} and \cite{tommasini}), it is remarked here the famous result that a vanishing damping has a strong detrimental effect on the flutter threshold, the so-called \lq Ziegler paradox' \citep{ziegler_0,B1956,KI2005,luongo}. A map of the real and imaginary parts of the pulsation $\omega$ as functions of the applied (dimensionless) load $p$ is provided in Fig.~\ref{fig16}, which refers to both the \lq ideal' case in which all dissipation sources are set to zero (on the left) and to the damped case in which both internal and external dissipation are present (on the right). The figure, relative to $\alpha=0.9819$, $v'(l)\bar{\chi} =0.092$, $\eta=0.348\cdot 10^{-3}$, $\gamma=50.764\cdot 10^{-3}$, confirms the fact that the flutter load is strongly decreased by dissipation, while the divergence load is slightly increased.
Preliminary results of experimental exploration of such an extension of the flutter domain has been announced quite recently by \cite{B2018} in the general context of dissipation-induced instabilities in reversible systems and spectral singularities associated with this effect \citep{OMN1996,KS2005}. It was emphasized by \cite{B2018} that the successful verification of the Ziegler paradox in viscoelastic structures moving in a resistive medium is strongly connected with the proper identification of all damping sources. In the following, the experiments are described in which the parameters governing flutter and divergence are varied in a way that the effect of damping can be assessed.

The behaviour of the Pfl\"uger's column was experimentally investigated covering a wide range of values of the ratio between the concentrated mass $M$ at the free end of the column  and the mass of the column $ml$, expressed through the parameter $\alpha$ ranging in the experiments between 0.7019 and 1.4256 for the eleven tested rods, see Table~\ref{table_beam}. The velocity of the belt was maintained constant during all the experiments and equal to 0.1\,m/s. To measure the critical load for flutter with high precision, the system was loaded by filling a container with water, so that an accuracy of one gram was achieved.

\begin{figure}[ht!]
  \begin{center}
    \includegraphics[width=\textwidth]{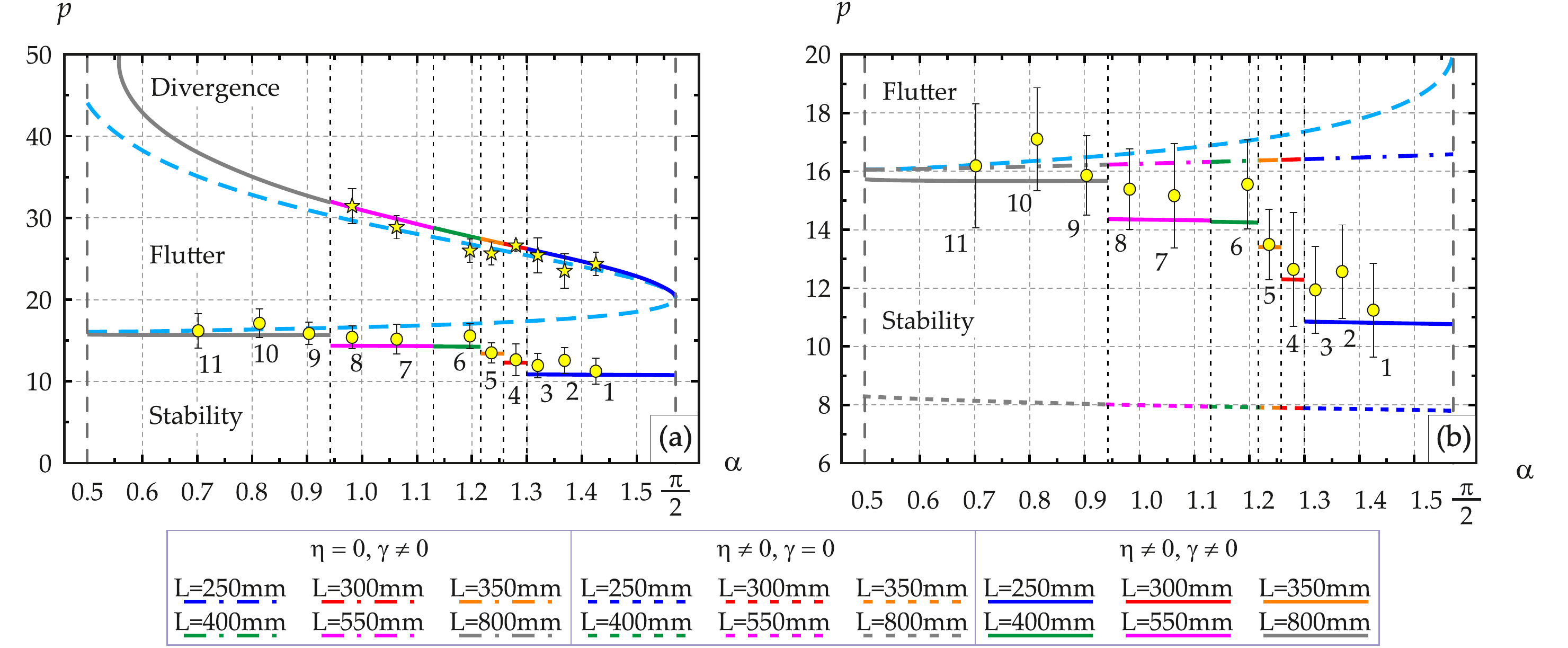}
    \caption{\footnotesize{(a)~Dimensionless critical load $p$ for flutter and divergence instability versus the mass ratio parameter $\alpha$. Experimental results are shown (spots and stars with error bars) together with theoretical predictions. The latter are reported for the ideal case (where damping is assumed to be absent) with a dashed curve and when both damping sources are present (solid curves). The different colors and the numbers identify the different samples tested (see Table~\ref{table_beam} for details). (b)~Detail of the flutter boundary, considering only internal damping (dotted curves), only external damping (dash-dotted curves) or both of them (solid curves). All the theoretical curves were computed considering that the load is not purely tangential ($v'(l)\bar{\chi}=0.092$). The experimental results confirm the decrease (the increase) of the critical load for flutter (for divergence) due to the effect of dissipation.
}}
    \label{fig15}
  \end{center}
\end{figure}

Before proceeding with the presentation of the results, notice that experiments were performed for rods of six different lengths, namely $L = \{250, 300, 350, 400, 550, 800\}$\,mm. 
Every sample has a different ratio of dimensionless coefficients of external and internal damping, so that according to the theory presented in \citep{tommasini,B2018} each ratio corresponds to a particular flutter boundary in the $p$ -- $\alpha$ plane of the dimensionless load $p$ and the mass-ratio parameter $\alpha$. In Fig.~\ref{fig15} only parts of the theoretical boundaries are shown that correspond to the same damping ratio as the samples marked by the symbols with the same color as the curve have. In the ideal case
(in which sources of damping are absent), the flutter boundary does not depend on the damping ratio \citep{kordas,LS2000,realff}. Theoretical predictions and experimental results are reported in Fig.~\ref{fig15}. In the figures, numbers identify the specific sample tested (the characteristics of the samples are reported in Table~\ref{table_beam}).

%
\begin{table}
\centering
\begin{tabular}{c|cccccccc}
\toprule[.5mm]
\multirow{ 2}{*}{Sample \#} & b  & h & L  & J & M  & $\alpha$ & $\eta,\times 10^{-3}$ & $\gamma,\times 10^{-3}$  \\
 & [mm] & [mm] & [mm]  & [mm$^4$] & [kg] & [-] & [-] & [-]  \\
\hline
1    & 1.90  & 24 & 250 & 13.718 & 0.105 & 1.426 & 1.059 & 24.705    \\
2    & 1.90  & 24 & 250 & 13.718 & 0.075 & 1.369 & 1.059 & 24.705    \\
3    & 1.90  & 24 & 250 & 13.718 & 0.060 & 1.320 & 1.059 & 24.705    \\
4    & 1.90  & 24 & 300 & 13.718 & 0.060 & 1.280 & 0.746 & 36.056    \\
5    & 1.92  & 24 & 350 & 14.156 & 0.060 & 1.236 & 0.557 & 48.368    \\
6    & 1.95  & 24 & 400 & 14.830 & 0.060 & 1.196 & 0.439 & 62.128    \\
7    & 2.98  & 24 & 550 & 52.927 & 0.089 & 1.063 & 0.348 & 50.764    \\
8    & 2.98  & 24 & 550 & 52.927 & 0.075 & 0.982 & 0.348 & 50.764    \\
9    & 3.07  & 24 & 800 & 57.869 & 0.089 & 0.903 & 0.177 & 102.538  \\
10  & 3.07  & 24 & 800 & 57.869 & 0.075 & 0.813 & 0.177 & 102.538  \\
11  & 3.07  & 24 & 800 & 57.869 & 0.060 & 0.702 & 0.177 & 102.538  \\
\toprule[.5mm]
\end{tabular}
\caption{Geometrical parameters and dimensionless coefficients for internal and external viscosity of the different structures tested with the flutter machine.}
\label{table_beam}
\end{table}

For conciseness, all the experimental data are reported in the figures, where \lq pieces' of the relevant theoretical curves have been plotted for comparison (marked with different colors in the figure). The ideal case is also included (dashed). Fig.~\ref{fig15}(b) is a detail of Fig.~\ref{fig15}(a) to highlight the flutter threshold, which is reported and compared to the theoretical curves relative to the presence of both damping sources and only one of them (either internal or external). It is visible that only when the two sources of dissipation act simultaneously the theoretical predictions fit correctly the experiments.

It has to be noted that the loads for the onset of flutter were measured simply by observing oscillations of the rod with deviations from rectilinearity of the structure, whereas the loads for the onset of divergence have been detected on the basis of the modes of vibrations, a circumstance which merits the following clarification.
The {\it linearized} equations of motion show that the instabilities of flutter and divergence are both characterized by an exponential growth in time of displacements (though the former is also accompanied by oscillations), which rapidly leads to a large amplitude motion and thus to a departure from the applicability of a linear theory. For the Ziegler double pendulum, both the experiments and the theoretical calculations show that the structure reaches a limit cycle when nonlinearities dominate \citep{noselli}. While our experiments clearly reveal the flutter instability threshold, the discrimination between flutter and divergence becomes difficult, because oscillations are experimentally found at every load beyond flutter and in all cases a limit cycle vibration is observed. This circumstance can be understood because quasi-static solutions are impossible for the Pfl\"uger's rod, so that an exponential growth should always \lq come to an end' for
a real system, in which displacements are necessarily limited. Therefore, oscillatory behaviour becomes in a sense necessary, and in fact was always observed in {\it both} the experiments and the numerical simulations (which will presented in subsection~\ref{strazmibal}).

As a conclusion, a discrimination between flutter and divergence cannot be simply based of the fact that oscillations do not occur for the latter case, which represents only the prediction of a linearized model. Divergence instability (never experimentally investigated before) is discriminated from flutter on the basis of the deformation of the rod at the onset of the instability, see Fig.~\ref{autovettori2}. Therefore, to detect divergence instability, high-speed videos at 240\,fps were taken (with a Sony high-speed camera, model PXW-FS5) and the motion of the rod at the beginning of  the instability monitored.

The experimental results obtained for flutter instability and shown in Fig.~\ref{fig15} lead to the important conclusion that {\it damping decreases the flutter load}, an effect that previously has never been experimentally documented in the known works of other groups. Damping has much less influence on the divergence load than on the flutter load, so that it is harder to conclude on this. Nevertheless, the experiments reported in Fig.~\ref{fig15} are in agreement with the conclusion that viscosity produces a shift of divergence towards higher loads.

\begin{figure}[ht!]
  \begin{center}
    \includegraphics[width=\textwidth]{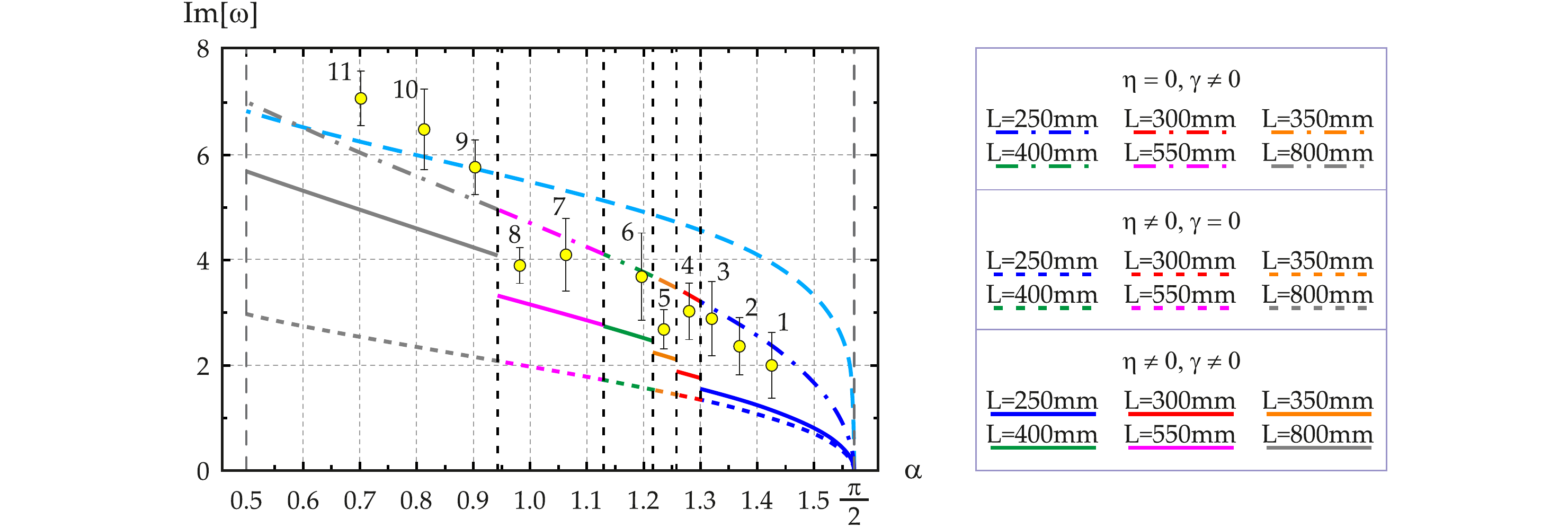}
   \caption{\footnotesize{Imaginary part of the dimensionless critical pulsations for flutter Im$[\omega]$ as a function of the mass ratio $\alpha$. Experiments are reported against different theoretical predictions. The latter are for the ideal case (without damping, dashed light blue curve), for the case with only internal (dotted curves) and only external (dot-dashed curves) damping and with both dampings (solid curves). The colors and numbers identify the different samples tested (see Table~\ref{table_beam} for details).  All the curves were computed considering that the load is not purely tangential ($v'(l)\bar{\chi}=0.092$).}}
    \label{fig17}
  \end{center}
\end{figure}

The variation of the dimensionless critical frequency for flutter with the mass ratio $\alpha$ (ranging between 0.5 and $\pi/2$) is reported in Fig.~\ref{fig17}. This behaviour was previously analyzed by \cite{SKK1976}, \cite{pedersen}, \cite{chen}, and \cite{ryu}, but only theoretically and without considering external and internal damping separately. Experiments, and therefore also theoretical predictions, are reported in Fig.~\ref{fig17} for different lengths of the rod, so that, while the theoretical predictions for the ideal case (without damping) are shown on the whole interval of $\alpha$  (dashed light blue), the cases where damping is present are illustrated by  fragments of the frequency curves occupying intervals of $\alpha$ that correspond to relevant samples. The situations in which either the internal or the external damping is considered separately correspond to the dotted and dot-dashed curves, respectively. The vertical bars and different colors denote the intervals corresponding to different lengths.

%
%
\subsection{Nonlinear dynamics of the Pfl\"uger's column under flutter and divergence}\label{strazmibal}

It has been already highlighted that, while the linearized theory predicts the exponential growth of divergence to occur at vanishing frequency, in reality limit cycle oscillations always occur. In fact, indefinite exponential growth is impossible in a real system and quasi-static solutions are impossible for the Pfl\"uger's rod. Hence, the discrimination between the two instabilities of flutter and divergence requires a numerical investigation in the full nonlinear range, which is presented below.

With the purpose of simulating the behaviour of the Pfl\"uger's column beyond the flutter threshold, a nonlinear computational model was devised and implemented in the commercial finite element software ABAQUS Standard~6.13-2. Specifically, 2-nodes linear elements of type B21 (in the ABAQUS nomenclature) were employed to discretize the viscoelastic rod of constant, rectangular cross section. A number of 20 elements was found to be sufficient to adequately resolve for the rod dynamics. A linear viscoelastic model of the Kelvin-Voigt type was implemented  for the constitutive response of the rod in a UMAT user subroutine, such that the bending moment $\mathcal{M}$ was proportional to the rod curvature and its time derivative through the elastic and viscous moduli, respectively. In the spirit of the linear model, the air drag (that is, the external source of damping) was accounted for with a distributed load, transversely applied to the rod and proportional to its velocity via the damping coefficient $K$. The follower forces $P_{tang}$ and $P_{orth}$ acting at the end of the rod (see the sketch of Fig.~\ref{fig2}) were set proportional to the virtually applied weight $W$ and to the respective friction coefficients $\mu_{tang}$ and $\mu_{orth}$, the former of which is defined by equation~(\ref{martins}), while the latter is defined by the same equation, but with the orthogonal velocity component $v_{orth}$ replacing $v_{tang}$ and with $\mu_0$ replaced by $0.09 \mu_0$.

From the computational standpoint, the friction law of Eq.~\eqref{martins} was implemented by exploiting the user subroutine UAMP together with the SENSOR functionality of ABAQUS.
All the dynamic analyses were performed by exploiting the default settings of ABAQUS Standard~6.13-2 and with a time increment of $10^{-4}$ s. The values of the geometric and material parameters were employed as summarized in Table~\ref{table_beam}. Finally, to check for the accuracy of the finite element model, the critical weight $W$ corresponding to the onset of flutter was numerically computed and compared with its theoretical value as provided by the linear model. Remarkable agreement was found for all the samples.

Three samples of different length were selected for the experimental and computational analysis of flutter and divergence in the nonlinear regime, namely sample 3, 5 and 8 of Table~\ref{table_beam}. These were tested at increasing values of $W = \{10,15,20,25,30,35,40\}$\,N so as to explore the nonlinear response of the system. While conducting the experiments, high-speed movies were recorded at 240\,fps. These were employed to track the position of the rod end in time, and consequently to estimate the frequency of the periodic oscillations reached at the limit cycle. Data acquired from an accelerometer mounted at free end of the rod were also exploited to the same purpose.
\begin{figure}[ht!]
  \begin{center}
    \includegraphics[width=.95\textwidth]{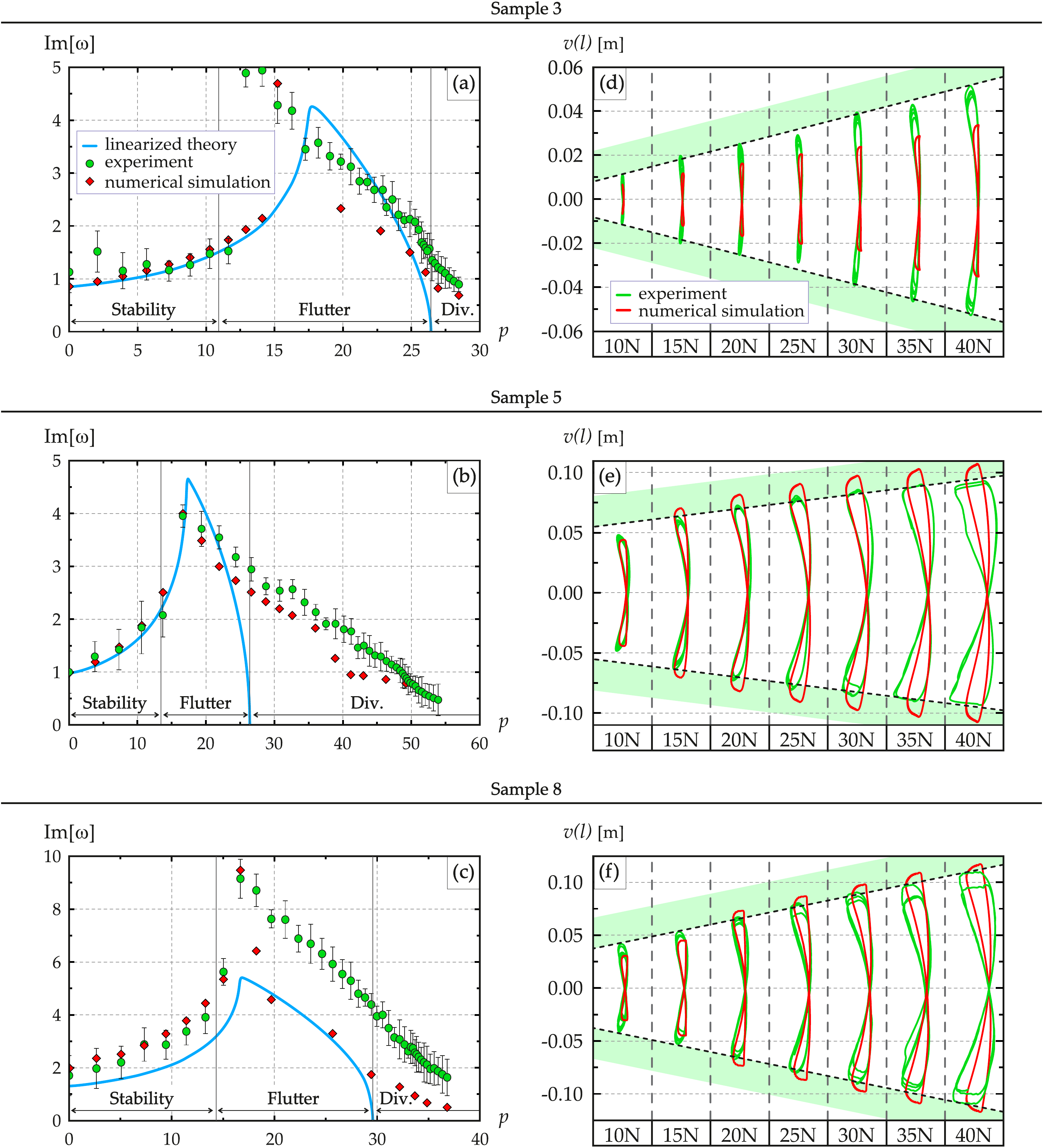}
    \caption{\footnotesize{(a)~The theoretical (from the linearized theory, blue solid curve), experimental (green spots) and computational (red diamonds) evolution of the pulsation Im[$\omega$] is reported for three different samples at increasing loads. (b)~Experimental (green) and numerical (red) trajectories of the Pfl\"uger's column end at increasing load. The velocity of the tape was set to 0.1\,m/s.
    }}
    \label{fig13}
  \end{center}
\end{figure}

Experimental results are reported in Fig.~\ref{fig13}, together with the predictions obtained from the linear model and from the numerical simulations. The frequency is reported versus the applied load in the left part of the figure, while the trajectories of the rod's end are reported on the right for the three tested rods.

The trajectories traced by the rod's end at different load levels, reported in Fig.~\ref{fig13} on the left, show an excellent agreement between numerical simulations and experimental observations and that the amplitude of the periodic motion linearly increases with the applied load $W$.

The evolution of the pulsation (and consequently of the frequency) shows an initial increase with the load level (in the flutter range), followed by a decrease (when divergence is approached).
While the linearized solution (solid blue curve) displays a well-defined divergence threshold, both the experimental results (green spots) and the numerical simulations (red diamonds) evidence that vanishing of the pulsation is attained only in an asymptotic sense, so that, as already remarked, divergence cannot be discriminated just looking at the absence of oscillations. At high load, when the divergence region is entered, the numerical solution of the fully nonlinear problem highlights that there is an initial exponential growth of the solution (as the linearized analysis predicts), but after this the rod reaches a maximum of flexure, stops for a moment and then shows a sort of \lq whipstroke-back', thus reaching an opposed inflected configuration. In other words, both the numerical solutions and the experiments show that the exponential growth predicted by the linearized solution degenerates into a sort of limit cycle oscillation.

\section{Conclusions}

A new experimental apparatus has been designed, manufactured and tested to investigate flutter and divergence instability in viscoelastic rods, as produced by tangentially follower forces. Experiments performed on a series of structures have determined the onset of flutter instability and, at higher load, the onset of divergence. The experiments have revealed that in both cases of flutter and divergence the system reaches a limit cycle oscillation, so that the structure/testing-machine complex provides an example of a self-oscillating system, which gains energy from a steady source and transforms this energy into a periodic motion with steady oscillation. Viscoelastic rod theory has been shown to correctly model the experimental results both in the linearized version, valid at small deflection, and in the nonlinear regime at high deformation. The first experimental determination has been provided of the destabilizing role of dissipation on the onset of flutter, whereas it has been shown that the same dissipation gives rise to an increase of the critical load for divergence. The experimental apparatus proposed in the present article allows for the application of follower forces to structures and therefore opens a new perspective in structural analysis and design.

\paragraph{Acknowledgements} O.K., D.M., G.N., M.T. gratefully acknowledge financial support from the ERC Advanced Grant \lq Instabilities and nonlocal multiscale modelling of materials' ERC-2013-ADG-340561-INSTABILITIES. D.B. thanks financial support from PRIN 2015 2015LYYXA8-006.


\appendix
\renewcommand\thesection{Appendix \Alph{section}}
\renewcommand\thesubsection{\Alph{section}.\arabic{subsection}}

\numberwithin{equation}{section}
\renewcommand{\theequation}{\Alph{section}.\arabic{equation}}

\section{Flutter and divergence instability for the Beck and Pfl\"uger's column}\label{derivation_equation}

The analysis of flutter and divergence for the Pfl\"uger's and Beck's columns is here continued from equations~(\ref{uno})--(\ref{sclero}). In particular, following the time-harmonic assumption of (\ref{sclero}), equation (\ref{checazzo}) yields a linear differential equation for $\tilde{v}(\xi)$, with the characteristic equation
\be{adim_diff_eq}
\lambda^4(1+\eta \omega)+\lambda^2 p + \gamma \omega+\omega^2 = 0,
\ee
which admits the solutions
\be{char_eq}
\lambda^2_{1,2}=\frac{\sqrt{p^2-4(1+\eta \omega)(\gamma \omega+\omega^2)} \mp p}{2(1+\eta \omega)}.
\ee
Therefore, the solution for $\tilde{v}$ can be written in the form
\be{hyperbolic_f}
\tilde{v}(\xi)=A_1 \sinh(\lambda_1 \xi) + A_2 \cosh(\lambda_1 \xi) + A_3 \sin (\lambda_2 \xi) + A_4 \cos(\lambda_2 \xi),
\ee
where $A_i$ ($i=1,..,4$) are arbitrary constants.

The boundary conditions of equation~(\ref{contorno}) can be rewritten in a dimensionless form as
\be{beckeq}
\begin{array}{lll}
\tilde{v}(0)=\tilde{v}'(0) = 0 & \mbox{at the clamped end,} \\[2.5mm]
\tilde{v}''(1) = 0 & \mbox{at the loaded end,} \\[2.5mm]
(1+\eta \omega)\tilde{v}'''(1) - (\chi-1)\tilde{v}'(1)p-\omega^2 \tan(\alpha) \tilde{v}(1) = 0 & \mbox{at the loaded end,}
\end{array}
\ee
where $\chi=1-\bar{\chi}$ represents the inclination of the applied dimensionless tangential force $p$. A substitution of the boundary conditions (\ref{beckeq}) in the solution (\ref{hyperbolic_f}) yields an algebraic system of equations which admits non-trivial solutions at the vanishing of the determinant of the matrix of coefficients
\begin{equation}
\label{matriciaccia}
\left[
\begin{array}{ccccc}
0 & 1 & 0 & 1 \\ [3 mm]
\lambda_1 & 0 & \lambda_2 & 0 \\ [3 mm]
\lambda_1^2 \sinh \lambda_1 & \lambda_1^2 \cosh \lambda_1  & -\lambda_2^2 \sin\lambda_2 & -\lambda_2^2 \cos\lambda_2 \\ [3 mm]
a_{41} & a_{42} & a_{43}   & a_{44}
\end{array}
\right]
\end{equation}
where
\begin{equation}
\label{soccia}
\begin{array}{ll}
a_{41} = (1+\eta \omega) \lambda_1^3 \cosh\lambda_1  - (\chi - 1) p \lambda_1 \cosh \lambda_1-\omega^2 \tan \alpha \sinh \lambda_1  ,   \\ [3 mm]
a_{42} = (1+\eta \omega)\lambda_1^3 \sinh\lambda_1   - (\chi - 1) p \lambda_1 \sinh \lambda_1-\omega^2 \tan \alpha \cosh \lambda_1 , \\ [3 mm]
a_{43} = -(1+\eta\omega)\lambda_2^3\cos\lambda_2  - (\chi - 1) p \lambda_2 \cos \lambda_2-\omega^2 \tan \alpha \sin \lambda_2 , \\ [3 mm]
a_{44} = (1+\eta \omega) \lambda_2^3 \sin\lambda_2  + (\chi - 1) p \lambda_2 \sin \lambda_2-\omega^2 \tan \alpha \cos \lambda_2 .
\end{array}
\end{equation}

Noting that the $\lambda_i$'s are functions of the applied load $p$, the pulsation $\omega$,  the viscosity $\eta$, and the external damping $\gamma$, the vanishing of the determinant of (\ref{matriciaccia}) corresponds to the frequency equation
\begin{equation}
\label{strasoccia}
f(p, \omega, \alpha, \gamma, \eta, \chi) = 0,
\end{equation}
which in explicit form reads
\be{trans_eq}
\begin{array}{rl}
f(p, \omega, \alpha, \gamma, \eta, \chi)&= \lambda_1\lambda_2(1+\eta \omega)(\lambda_1^4+\lambda_2^4)+\lambda_1\lambda_2 p(\chi-1)(\lambda_2^2-\lambda_1^2) + \\[3.5 mm]
&+\lambda_1\lambda_2[2(1+\eta \omega)\lambda_1^2\lambda_2^2-p(\chi-1)(\lambda_2^2-\lambda_1^2)]\cosh \lambda_1 \cos \lambda_2 + \\[3.5 mm]
&+\lambda_1^2 \lambda_2^2[2p(\chi-1)+(1+\eta \omega)(\lambda_2^2-\lambda_1^2)]\sinh \lambda_1\sin \lambda_2 + \\[3.5 mm]
&\displaystyle -\omega^2 \tan \alpha (\lambda_1^2+\lambda_2^2)[\lambda_2\sinh\lambda_1\cos\lambda_2-\lambda_1\cosh\lambda_1\sin\lambda_2].
\end{array}
\ee

For given values of the parameters $\chi$, $\eta$, and $\gamma$ (which were identified from experiments), the transcendental equation (\ref{strasoccia}) describes the boundaries for which flutter occurs and for which divergence occurs. In particular, the system is unstable by divergence when the pulsation $\omega$ is real and positive, while flutter instability occurs for complex $\omega$ with Re[$\omega$]$>$0.

\begin{figure}[ht!]
  \begin{center}
    \includegraphics[width=\textwidth]{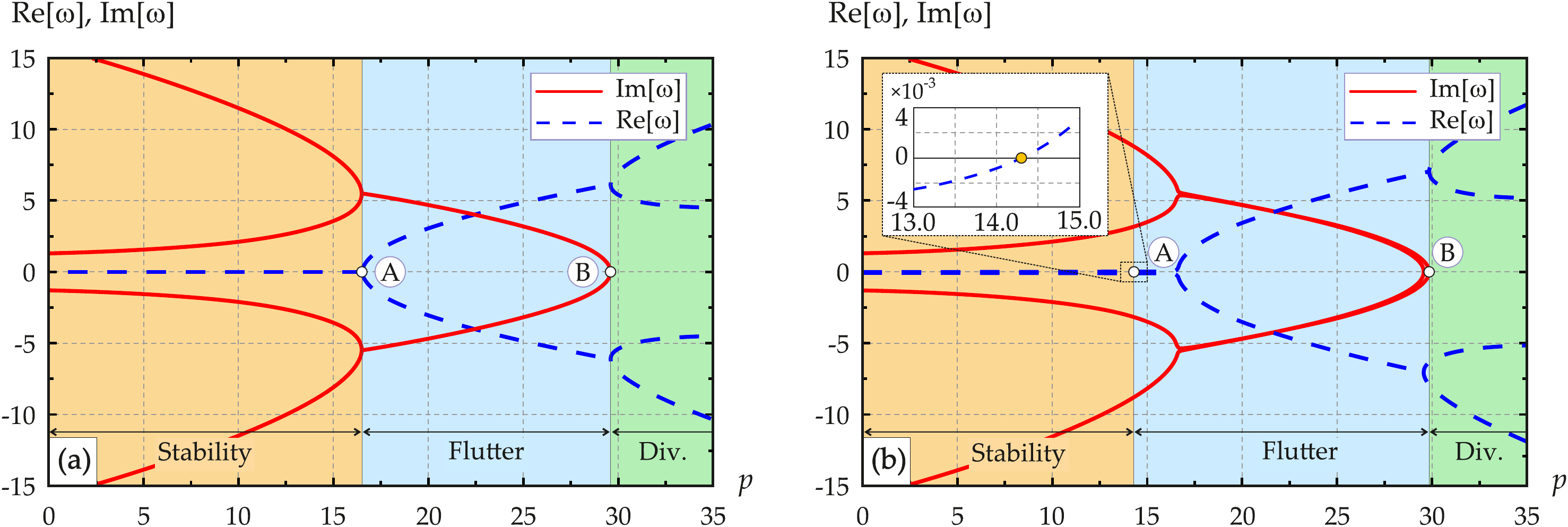}
    \caption{\footnotesize{Branches of the real Re[$\omega$] and imaginary Im[$\omega$] parts of the pulsation $\omega$ defining the vibration of the Pfl\"uger's column, with a mass ratio $\alpha = 0.9819$ and an inclination of the end force $v'(l)\bar{\chi}=0.092$. The ideal case (absence of damping) is reported on the left (a), where flutter (marked with the letter A) occurs at $p=16.499$ and divergence (marked with the letter B) at $p=29.597$. The case in which both the external and internal dampings are present (with coefficients corresponding to our experimental setup, i.e.
$\eta=0.348\cdot 10^{-3}$ and $\gamma=50.764\cdot 10^{-3}$) is shown on the right (b). Here, the flutter load decreases to $p=14.318$, while the divergence load increases to $p=29.575$. Flutter occurs when a real branch of the pulsation $\omega$ becomes positive (with non zero values of the imaginary part of the pulsation).}}
    \label{fig16}
  \end{center}
\end{figure}

In the following the case in which instability is obtained under the initial hypothesis of null damping is referred to as \lq the ideal case'. Generally speaking, damping decreases the flutter loads and increases the divergence load, an effect which can be verified from Fig.~\ref{fig16}, where the real and imaginary parts of the pulsation of the system are reported for the ideal (a) and damped (b) cases, so that it becomes evident that damping enlarges the flutter domain. This effect is more pronounced when internal and external dampings are both increased.

\renewcommand\thesection{Appendix \Alph{section}}
\renewcommand\thesubsection{\Alph{section}.\arabic{subsection}}

\numberwithin{equation}{section}
\renewcommand{\theequation}{\Alph{section}.\arabic{equation}}

\section{Identification of the internal and external damping coefficients}\label{identification}

In the present study, two sources of damping have been accounted for in equation (\ref{eq_diff}), a damping internal to the rod via parameter $E^*$ of equation (\ref{uno}), and another due to the air drag via parameter $K$ in the equations of motion (\ref{eq_diff}). Often these two damping sources are condensed in a single coefficient, see \cite{chopra} and \cite{clough_penzien}, but it is well-known that in problems of flutter a careful distinction has to be maintained between the different sources of damping \citep{SW1975,detinko}.

\begin{figure}[ht!]
  \begin{center}
    \includegraphics[width=\textwidth]{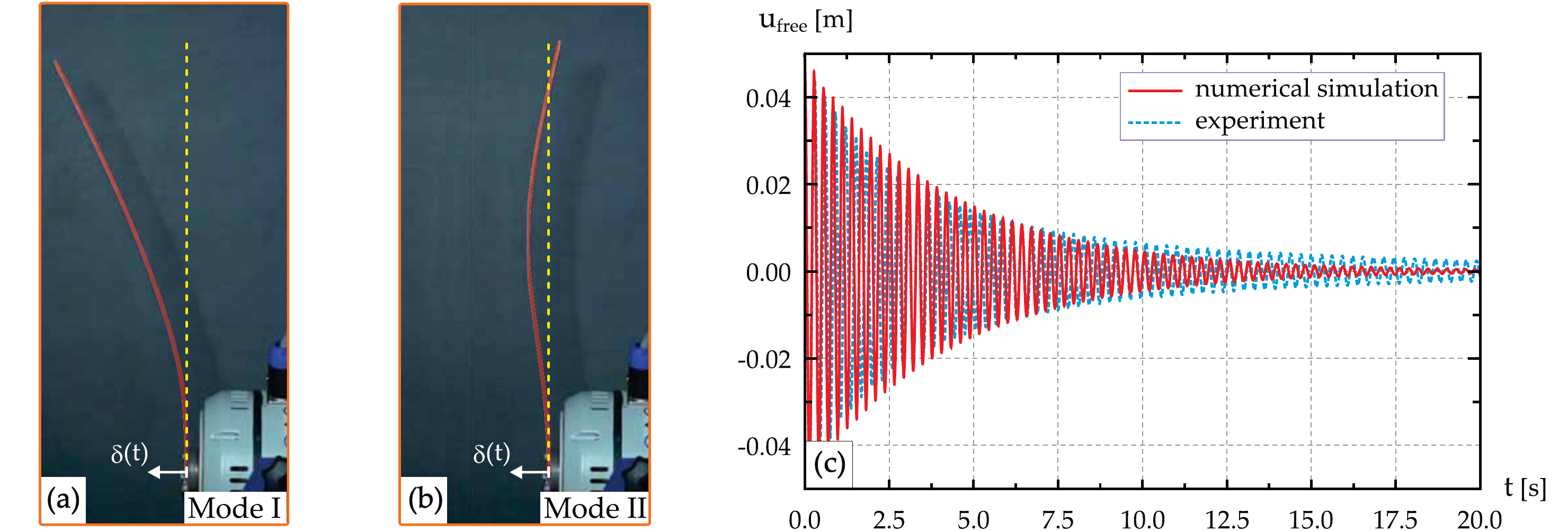}
    \caption{\footnotesize{Identification of the internal and external damping coefficients for an oscillating rod. (a)~The experimental setup showing a rod mounted on a shaker in a cantilever configuration, which vibrates in the first and second resonant mode while a sinusoidal displacement $\delta(t)$ of amplitude 10\,mm is applied at the clamp. Using this setup, after a steady state regime was reached, the shaker was turned off and the oscillations of the free end were monitored. With the data acquired, and using a modified logarithmic decrement approach, two damping ratios $\zeta$ were identified. Eventually the damping coefficients $E^*$ and $K$ were calculated. (b)~These coefficients have been eventually validated by imposing and suddently releasing a displacement of 50\,mm at the end of the rod in a cantilever configuration. The subsequent motion of the rod end was recorded with a high-speed camera and tracked with a software developed using Mathematica. The outcome of the experiment was compared with the simulations obtained with a model of the rod implemented in ABAQUS Standard~6.13-2. The good agreement between the experiment and the simulation validates the adopted identification procedure.
}}
    \label{fig4}
  \end{center}
\end{figure}

Therefore, experiments have been performed to identify the viscous modulus $E^*$ which accounts for the internal damping and the coefficient $K$ parametrizing the external damping due to air drag. To this purpose, the elastic rod used for the flutter experiments was mounted on a shaker (Tira vib 51144, frequency range 2-6500\,Hz, equipped with a Power Amplifier Tira BAA 1000) in a cantilever configuration. On the free end of the rod an accelerometer (352A24 from PCB Piezotronics) was positioned and a sinusoidal base displacement $\delta(t)$, with a maximum amplitude of 10\,mm, was imposed at the clamped end of the rod. The shaker was tuned to the first and second resonance mode, as shown in Fig.~\ref{fig4}, until a steady state was reached. Subsequently, the shaker was turned off and the oscillations of the free end of the cantilever rod were monitored. The data acquired from the accelerometer mounted on the free end were employed in a modified logarithmic decrement approach (details on this method are reported in~\ref{app_A}) in order to obtain two distinct damping ratios $\zeta$. Eventually the following values were computed for the internal and external damping coefficients, namely $E^*=2.139796\cdot 10^6\,kg\, m^{-1} s^{-1}$ and $K=1.75239\cdot10^{-5}\,kg\, m^{-1} s^{-1}$, respectively.

The identified damping parameters were eventually validated through the following additional experiment: a displacement of 50\,mm was imposed at the end of the viscoelastic rod, in a cantilever configuration, with the aid of a wire. The motion of the rod arising from the sudden cut of the wire was recorded with a high-speed camera (Sony high-speed camera, model PXW-FS5). A circular marker, applied at the end of the beam, was tracked frame by frame with an appropriate code implemented in Mathematica. The motion of the rod was then simulated with a computational model implemented in ABAQUS Standard~6.13-2, assuming the two previously evaluated damping parameters. The experimental and the computational results (reported in light blue and red in Fig.~\ref{fig4}, respectively) were found in good agreement, so that the estimated damping parameters were successfully validated.


\renewcommand\thesection{Appendix \Alph{section}}
\renewcommand\thesubsection{\Alph{section}.\arabic{subsection}}

\numberwithin{equation}{section}
\renewcommand{\theequation}{\Alph{section}.\arabic{equation}}

\section{Modified logarithmic decrement approach to identify the external and internal damping coefficients}
\label{app_A}

The scope of the present Section is to detail the identification technique employed to evaluate the external and internal damping coefficients. The free vibrations of a viscoelastic rod is governed by equation~(\ref{eq_diff}) with $P=0$, which can be rewritten as
\be{eq_gen}
EJ v''''+E^* J \dot{v}''''+K \dot{v}+m \ddot{v}=0.
\ee

Before analyzing the case in which the clamped end of the rod is subject to a sinusoidal excitation of pulsation $\bar{\omega}$ (which will be relevant for the identification of the damping coefficients), free vibrations are considered of the cantilever configuration with fixed clamp.

\subsection{Free vibration of a cantilever rod} \label{Section2}

Free vibrations of a viscoelastic cantilever rod are analyzed assuming time-harmonic solutions in the form $v(x,t)=Y(x)\exp(-i \omega t)$ [notice the slight difference with respect to the representation of equation~(\ref{sclero})], yielding
\be{damped_Y}
\sum\limits_{n=1}^\infty Y_n^{IV}-\Lambda_n^4 Y_n=0 ,
\ee
where the coefficients $\Lambda_n^4$ are real quantities defined as
\be{}
\Lambda_n^4=\frac{m \omega_n^2+i \omega_n K}{EJ-i \omega_n E^* J}.
\ee
The solution of Eq.~(\ref{damped_Y}) under the time-harmonic assumption can be represented as
\be{free_coef}
Y(x)=\sum\limits_{n=1}^\infty Y_n(x)=\sum\limits_{n=1}^\infty C_{1,n}\sin \Lambda_n x + C_{2,n}\cos \Lambda_n x + C_{3,n}\sinh \Lambda_n x + C_{4,n}\cosh \Lambda_n x ,
\ee
where the constants $C_{i,n}$ depend on the boundary conditions, which for a cantilever rod read
\be{bcbeck}
Y(0)=Y'(0)=Y''(l)=Y'''(l)=0.
\ee

A substitution of equations~(\ref{bcbeck}) in the representation (\ref{free_coef}) leads to
\be{sistema_1bc}
 \left[ \begin{array}{cccc}
0 & 1 & 0 & 1 \\ [2.0mm]
\Lambda_n & 0 & \Lambda_n & 0 \\ [2.0mm]
-\Lambda_n^2 \sin \Lambda_n l & -\Lambda_n^2 \cos \Lambda_n l & \Lambda_n^2 \sinh \Lambda_n l & \Lambda_n^2 \cosh \Lambda_n l \\ [2.0mm]
-\Lambda_n^3 \cos \Lambda_n l & \Lambda_n^3 \sin \Lambda_n l & \Lambda_n^3 \cosh \Lambda_n l & \Lambda_n^3 \sinh \Lambda_n l
\end{array} \right]
\left(
\begin{array}{c}
C_{1,n} \\ [2.0mm]
C_{2,n} \\ [2.0mm]
C_{3,n} \\ [2.0mm]
C_{4,n}
\end{array} \right)
=
\left(
\begin{array}{c}
0 \\ [2.0mm]
0 \\ [2.0mm]
0 \\ [2.0mm]
0
\end{array} \right)
.
\ee
The first two equations yield $C_{4,n}=-C_{2,n}$ and $C_{3,n}=-C_{1,n}$, so that the algebraic system reduces to
\be{sistema_1bc_red}
 \left[ \begin{array}{cc}
\Lambda_n^2 (\sin \Lambda_n l +\sinh \Lambda_n l)& \Lambda_n^2 (\cos \Lambda_n l + \cosh \Lambda_n l) \\ [2.0mm]
\Lambda_n^3 (\cos \Lambda_n l +\cosh \Lambda_n l)& \Lambda_n^3 (-\sin \Lambda_n l + \sinh \Lambda_n l)
\end{array} \right]
\left(
\begin{array}{c}
C_{1,n} \\ [2.0mm]
C_{2,n}
\end{array} \right)
=
\left(
\begin{array}{c}
0 \\ [2.0mm]
0
\end{array} \right)
,
\ee
and imposing the determinant of the matrix to vanish provides
\be{trasc}
\cos \Lambda_n l \cosh \Lambda_n l=-1.
\ee
The just obtained transcendental equation defines the values
\be{}
\begin{array}{cccc}
\Lambda_1 l=1.875..., & \Lambda_2 l=4.694..., & \Lambda_n l=\frac{\pi}{2}(2n-1). \\
\end{array}
\ee
Now, the solution (\ref{free_coef}) can be expressed in terms of one arbitrary constant $C_{2,n}$, such that
\be{}
C_{1,n}=-\frac{\cos \Lambda_n l + \cosh \Lambda_n l} {\sin \Lambda_n l+\sinh \Lambda_n l}C_{2,n},
\ee
which leads to the general solution for the free vibrations of a cantilever rod in the form
\be{free_undamped}
v(t,x)=
\sum\limits_{n=1}^\infty  C_{2,n} Y_n(x) e^{-\omega_n^I t} \left(\cos \omega_n^R t +\frac{\omega_n^I}{\omega_n^R}\sin\omega_n^R t \right) ,
\ee
where the pulsation is split into the real $\omega_n^R$ and imaginary $\omega_n^I$ part, so that $\omega_n=\omega_n^R+i\,\omega_n^I$, and
\be{ipsilon}
Y_n(x) =
\cos \Lambda_n x-\cosh \Lambda_n x  -\frac{\cos \Lambda_n l+\cosh \Lambda_n l}{\sin \Lambda_n l + \sinh \Lambda_n l} \left(
\sin \Lambda_n x -\sinh \Lambda_n x\right) .
\ee

Notice that the free vibration solutions $Y_n(x)$ satisfy the orthogonality relations
\be{}
\begin{array}{lcr}
\int_0^l Y_n(x)Y_h(x)dx=0 & \text{for}\,\, h\neq n
\end{array}
\ee
and, because of Eq.~(\ref{damped_Y}), the property
\be{}
Y^{IV}_n(x)=\Lambda_n^4 Y_n(x).
\ee

The following quantities, which will be useful later, are also introduced
\be{}
\begin{array}{lr}
\Gamma_n:=\int_0^l Y^2_n(x)dx, &~~ \Gamma_n \Lambda_n^4:=\int_0^l Y^{IV}_n(x)Y_n(x)dx .
\end{array}
\ee


\subsection{Solution for a cantilever rod subject to sinusoidal base motion}
\label{Section4}

A sinusoidal displacement of amplitude $U_0$, namely
\begin{equation}
\label{straz}
\delta(t)=U_0 \sin\bar{\omega}t,
\end{equation}
is prescribed at the clamp $x=0$ of a viscoelastic rod in a cantilever configuration, in the presence of both internal and external damping. Under these conditions, the solution $u(x,t)$ can be written as the sum of a flexural displacement (function of space and time) $v(x,t)$ and the rigid-body motion (\ref{straz})
\begin{equation}
\label{gnoccona}
u(x,t) = v(x,t) + \delta(t),
\end{equation}
so that $v(0,t)=0$. A substitution of equation~(\ref{gnoccona}) in the differential equation of motion~(\ref{eq_gen}) by neglecting the term $\dot{\delta}(t)$ yields a differential equation for $v(x,t)$, that is
\be{eq_gen_no}
EJ v''''+E^* J \dot{v}''''+K \dot{v}+m \ddot{v}= m U_0 \bar{\omega}^2 \sin \bar{\omega}t .
\ee
Therefore, the effect of the movement at the clamp can be considered as the effect of a diffused load $f(t) = m U_0 \bar{\omega}^2 \sin \bar{\omega}t$, defined per unit length of the rod.

The solution of equation~(\ref{eq_gen_no}) is the sum of the solution of the associated homogeneous equation (which governs the free vibrations of the rod) and a particular solution $v_p(x,t)$.
The latter solution can be sought in the form
\begin{equation}
\label{trac}
v_p(x,t)=\sum\limits_{n=1}^\infty Y_n(x) y_n(t),
\end{equation}
where the modal functions $Y_n(x)$ are defined by equation~(\ref{ipsilon}) and the $y_n(t)$ are for the moment unknown. A substitution of equation~(\ref{trac}) into equation~(\ref{eq_gen_no}) yields
\be{general}
\sum\limits_{n=1}^\infty Y_n''''(x)y_n(t)+\frac{E^*}{E}Y_n''''(x)\dot{y}_n(t)+\frac{K}{EJ}Y_n(x)\dot{y}_n(t)+\frac{m}{EJ}Y_n(x)\ddot{y}_n(t)=\frac{f(t)}{EJ}.
\ee

A multiplication of the previous equation by $Y_h(x)$ and an integration over the length of the rod $l$ yields
\be{general_simp}
\Gamma_n \Lambda_n^4 y_n(t)+\Gamma_n \left( \frac{K}{EJ}+\frac{E^*}{E}\Lambda_n^4\right)\dot{y}_n(t)+\Gamma_n \frac{m}{EJ}\ddot{y}_n(t)=F_n \frac{f(t)}{EJ},
\ee
where $F_n=\int_0^l Y_n(x)dx$. Eq.~(\ref{general_simp}) is formally identical to the equation of motion which governs the oscillations of a single-degree-of-freedom system with a mass $m_n$, a damper of constant $c_n$ and a spring of stiffness $k_n$
\be{sdof}
m_n \ddot{y}_n(t)+c_n \dot{y}_n(t)+k_n y_n(t)=p_n \sin \bar{\omega} t,
\ee
where
\begin{equation}
m_n=\Gamma_n\frac{m}{EJ}, ~~~~ c_n=\Gamma_n\left(\frac{K}{EJ}+\frac{E^*}{E}\Lambda_n^4 \right) , ~~~~ k_n=\Gamma_n \Lambda_n^4 ,  ~~~~ p_n=F_n \frac{m U_0 \bar{\omega}^2}{EJ}.
\end{equation}

It is expedient to rewrite equation~(\ref{sdof}) in the form
\be{sdof_simp}
\ddot{y}_n(t)+2\alpha_n \zeta_n \dot{y}_n(t)+\alpha_n^2 y_n(t)=a_n \sin \bar{\omega}t
\ee
where
\be{sdof_coeff}
\alpha_n^2=\frac{k_n}{m_n}=\frac{EJ}{m}\Lambda_n^4, ~~~~ 2\alpha_n \zeta_n=\frac{c_n}{m_n}=\frac{K}{m}+\frac{E^* J}{m}\Lambda_n^4, ~~~~
a_n=\frac{p_n}{m_n}= \frac{F_n}{\Gamma_n} U_0 \bar{\omega}^2.
\ee

The solution of the differential equation~(\ref{sdof_simp}) consists of the sum of the solution of the associated homogeneous equation and of a particular integral, which can be found in the form
\begin{equation}
\label{cogliorotto}
y_{n,part}(t)=A_n \sin \bar{\omega}t+B_n\cos \bar{\omega}t,
\end{equation}
yielding
\be{a_b_part}
\begin{array}{cc}
A_n=a_n \left[1-\left(\frac{\bar{\omega}}{\alpha_n} \right)^2 \right] N_n ,& B_n=-2 a_n \zeta_n  \left(\frac{\bar{\omega}}{\alpha_n} \right) N_n ,
\end{array}
\ee
where $N_n$ is the so called dynamic amplification factor
\be{daf}
N_n(\alpha_n,\zeta_n)=\frac{1}{\left[1-\left(\frac{\bar{\omega}}{\alpha_n} \right)^2 \right]^2+\left[2\zeta_n \frac{\bar{\omega}}{\alpha_n} \right]^2}.
\ee

The solution of the homogeneous equation associated to equation~(\ref{sdof_simp}) is
\begin{equation}
\label{rottocoglio}
y_{n,hom}(t)=\exp(-\alpha_n \zeta_n t)\left( C_n \sin \alpha_{n,d} t+D_n \cos \alpha_{n,d}t \right) ,
\end{equation}
where $\alpha_{n,d}=\alpha_n \sqrt{1-\zeta_n^2}$ are the damped pulsations of the system. As regards the coefficients $C_n$ and $D_n$, these can be found by imposing the initial conditions
\be{}
\begin{array}{cc}
y_{n}(0)=X_0, & \dot{y}_{n}(0)=V_0
\end{array}
\ee
where
\be{}
y_{n}(t)=y_{n,hom}(t)+y_{n,part}(t),
\ee
leading to the expressions
\be{}
\begin{array}{cc}
\displaystyle C_n=\frac{1}{\alpha_{n,d}}\left[X_0 \alpha_n \zeta_n+V_0+a_n \bar{\omega} N_n\left(\frac{\bar{\omega}^2}{\alpha_n^2}+2\zeta_n^2-1 \right) \right] ,& D_n=X_0+2 a_n \zeta_n\frac{\bar{\omega}}{\alpha_n}N_n.
\end{array}
\ee


\subsection{Identification procedure}
\label{Section5}

The identification procedure is based on experiments in which the cantilevered rod is set in a steady oscillation at the $n$-th mode through excitation with a sinusoidal base motion. Starting from this situation, the base motion is stopped and the subsequent decaying oscillations are monitored. This transient motion is governed by equation~(\ref{rottocoglio}), in which the sinusoidal part has a period $T = 2\pi/\alpha_{n,d}$.
If the transient motion exhibits a peak of displacement at $t=t^*$, other peaks will occur at every cycle, even after $s$ cycles.
Therefore, the logarithmic decrement is defined as
\begin{equation}
\delta_s = \log \frac{y_{n,hom}(t^*)}{y_{n,hom}(t^* + 2\pi s/\alpha_{n,d})},
\end{equation}
a quantity which can be measured and satisfies the relation
\be{log_decay}
\zeta_n=\frac{\delta_s}{2\pi s \alpha_n/\alpha_{n,d}}\approx \frac{\delta_s}{2\pi s}.
\ee
From equation (\ref{sdof_coeff})$_2$ the following identity is finally obtained
\begin{equation}
\label{caaaz}
\frac{1}{\Lambda_n^2} \left(\frac{K}{J}+E^* \Lambda_n^4 \right)\sqrt{\frac{J}{m E}} = \frac{\delta_s}{\pi s},
\end{equation}
so that if $\delta_s$ is measured at cycle $s$ for two modes of vibration ($n=1$ and $n=2$ have been used in our experiments) and $J$, $E$ and $m$ are known from independent evaluations,
(\ref{caaaz}) provides two equations for the two unknown damping coefficients $E^*$ and $K$. Note that the logarithmic decrement can be equivalently measured as the ratio between peak displacements or between accelerations, because the latter are proportional to the former through a constant.

\end{document}